# THE PROBLEM OF THE COSMOLOGICAL CONSTANT:
EMQG and the Recent, High Red-Shift, Supernovae Observational Evidence for a Positive Cosmological Constant

**Tom Ostoma and Mike Trushyk**
Email: emgg@rogerswave.ca
Tuesday, March 09, 1999

**ABSTRACT**

ElectroMagnetic Quantum Gravity (EMQG) is applied to the problem of the Cosmological Constant. EMQG is a quantum gravity theory (ref. 1) in which the virtual particles of the quantum vacuum play a very important role in all gravitational interactions, and also in accelerated motion. According to general relativity, the cosmological constant can be thought of as the measure of the mass-energy density contained in empty space alone. According to EMQG theory (and quantum field theory in general), empty space is populated by vast numbers of virtual particles, consisting of virtual fermion and virtual anti-fermion particles, which posses mass, and also virtual boson particles of all the various force particle species. Therefore the problem of the cosmological constant is essentially equivalent to a determination of the mass contributed by all the virtual particles of the vacuum to the overall curvature and dynamics of the entire universe.

Originally we proposed a cosmological constant value that was essentially zero, because of the symmetrical creation of virtual particle/anti-particle pairs in the vacuum. We argued that anti-matter, which consists of half the virtual fermion particles of the quantum vacuum, is associated with a negative gravitational force such that the attractive gravitational forces of virtual fermion particles are exactly balanced by the repulsive gravitational forces of virtual anti-fermions (which violates the equivalence principle, an experimentally verifiable conclusion of EMQG theory). The virtual bosons are handled as a separate issue in EMQG, and do not contribute to the cosmological constant. A value of exactly zero for the cosmological constant seems to contradict the recent high red shift, type I(a) supernovae observations, which report a positive cosmological constant, and therefore <u>*an accelerated cosmic expansion*</u>. Our original analysis was based on the assumption of perfect symmetry in the creation and destruction of virtual fermion and virtual anti-fermion particle pairs in the quantum vacuum, which is in accordance with the existing laws of conservation of both electric charge, and also of the new law of conservation of gravitational 'mass charge'. We now believe that this may not be necessarily true, and that EMQG can allow the possibility of a very tiny value for the cosmological constant. Furthermore, it turns out that this question can only be resolved when a solution to the problem of baryon asymmetry in the universe is resolved. In other words an answer must be found as to why our universe seems to contain only real matter, and very little, if any anti-matter.

1. **INTRODUCTION**

*"The cosmological constant is probably the quantity in physics that is most accurately measured to be zero: observations of departures from the Hubble law for distant galaxies place an upper limit of the order of $10^{-120}$."*

- Stephen Hawking (1984)

*"A new component of the Universe which leads to an accelerated cosmic expansion is found from the measurements of distances to high-redshift type Ia supernovae. …. A positive cosmological constant is inferred from these measurements."*

- B. Leibundgut, G. Contardo, P. Woudt, J. Spyromilio (1998)

Einstein originally introduced the cosmological constant ($\Lambda$) in 1917. It was a *'fudge factor'* that he added to his gravitational field equations to make the universe remain in the steady state or non-expanding state, as was generally believed in his time. This new term corresponded to a non-zero energy-momentum tensor of the vacuum in his gravitational field equations. Without this cosmological constant term the only stable cosmological solution to his gravitational field equations is for an expanding universe. Einstein later abandoned the cosmological constant when E. Hubble's observations in the 1920's confirmed that the universe was actually expanding. In fact, Einstein later considered the addition of $\Lambda$ to his field equations as the *biggest* blunder of his life.

The cosmological constant basically represents the measure of the mass-energy density contained in empty space alone according to general relativity. From the perspective of classical physics, the vacuum is empty by definition, and therefore $\Lambda$ should be equal to zero. From the perspective of quantum physics, the vacuum is far from empty. By some estimates, the quantum vacuum contains a staggering $10^{90}$ virtual particles per cubic meter at any given instant of time. Furthermore, for every type of a virtual particle that contains mass (virtual electrons, virtual quarks, etc.) there exists a corresponding variety of virtual anti-particle (virtual anti-electrons, etc.). The virtual photons and virtual gravitons are their own anti-particle. That is a lot of hidden mass that should be contributing something to the gravitational dynamics of the universe, and to the overall curvature of space-time.

Thus, there is plenty of contributions from the virtual particles of the quantum vacuum to the total mass of the universe. If this is true, why is their presence not felt gravitationally? It has been said that the cosmological constant is **the most** striking problem in contemporary fundamental physics (ref. 7). In fact, theoretical predictions from quantum field theory for the cosmological constant differs from observation by at least a factor of $10^{45}$ and possibly, by a staggering $10^{120}$! Experimentally, it is safe to say that its actual value of $\Lambda$ is very close to zero (even bearing in mind the latest supernovae observations). In fact, S. Hawking (ref. 7) once stated that "**the cosmological constant is probably the quantity in physics that is most accurately measured to be zero**" as in his quote at the beginning of this section.

Recently, the value of the cosmological constant has come into question. A group of physicists and astronomers have recently conducted a detailed observational program to determine the dynamics of distant galaxies through supernovae explosions. They have found evidence from their measurements of distances to high-redshift type I(a) supernovae in distant galaxies that the universe has an accelerated cosmic expansion. It can be shown from Einstein's gravitational field equations that the addition of a small positive cosmological constant term can account for this. We will use a new quantum gravity theory to look at the problem of the cosmological constant from the perspective of quantum field theory.

## 2. INTRODUCTION TO EMQG AND THE COSMOLOGICAL CONSTANT

*" It always bothers me that, according to the laws as we understand them today, it takes a computing machine an infinite number of logical operations to figure out what goes on in no matter how tiny a region of space and no matter how tiny a region of time. .... why should it take an infinite amount of logic to figure out what one tiny piece of space-time is going to do?"*

*- Richard Feynman*

We have developed a quantum theory of gravity called ElectroMagnetic Quantum Gravity or EMQG (ref. 1). Originally we were motivated to provide a quantum gravity theory that is manifestly compatible with a Cellular Automata (CA) model (ref. 2), and also a theory based a pure particle exchange process for all forces, with gravity being no exception. This is in general accordance to the principles of quantum field theory. The particle exchange paradigm fits very well within the CA model.

As a result of our investigation into a theory of quantum gravity that is compatible with CA theory, we discovered the hidden quantum processes behind the Einstein Equivalence principle (ref. 1) and the quantum origins of Newtonian Inertia (ref. 3). It turns out the virtual particles of the quantum vacuum is the missing link in quantizing gravity, and plays an *essential role* in gravity, as it does for inertia. One of the key discoveries of EMQG theory is the existence of 'negative gravitational mass charge' for all anti-particles, and the great similarity between the photon and graviton particles, which are the mediators of the electromagnetic and gravitational forces, respectively. It turns out that a repulsive gravitational force exists between any two anti-particles, in clear violation of the Einstein Weak Equivalence Principle (WEP). The Equivalence principle has been tested to phenomenal accuracy since it's discovery by Einstein in the early 1900's. However, it has not been tested very well for anti-matter, which still cannot be manufactured in bulk quantities. The behavior of gravitational interactions of bulk anti-matter is totally unknown experimentally. EMQG predicts that the equivalence principle totally breaks down for anti-matter, and is only approximately true for ordinary matter. If a very large and incredibly small mass are dropped on the surface of the earth, the larger mass will arrive slightly earlier then the small mass, in clear violation of the equivalence principle!

We have found a deep connection between EMQG and the problem of the cosmological constant, because virtual particles of the quantum vacuum plays a dominate role in gravitational interactions, and in accelerated frames. The problem of the cosmological constant is equivalent to asking what relationship exists between the vast numbers of virtual particles and virtual anti-particles in the quantum vacuum with the total mass, and curvature of the entire universe. Since the virtual fermions of the quantum vacuum roughly consist of an equal mix of matter and anti-matter types with equal and opposite gravitational mass, the vacuum is balanced and contributes very little to the overall dynamics of the universe.

In formulating a successful quantum gravity theory, a mechanism must be found that produces the gravitational force, while somehow linking to the principle of equivalence of inertial and gravitational mass. In addition, this mechanism should naturally lead to 4D space-time curvature and should also be compatible with the principles of general relativity theory. Furthermore, it should be compatible with the general principles of quantum field theory. EMQG theory meets all these criteria. EMQG provides a new understanding of gravitation, and is also a testable theory, since it predicts new experimental results that cannot be explained by conventional general relativity theory.

Nature has another long-range force called electromagnetism, which has been described successfully by the principles of quantum field theory. This well known theory is called Quantum ElectroDynamics (QED), and this theory has been tested for electromagnetic phenomena to an unprecedented accuracy. It is therefore reasonable to assume that gravitational force should be a similar process, since gravitation is also a long-range force like electromagnetism. However, a few obstacles lie in the way that complicate this line of reasoning:

(1) **Gravitational force is observed to be <u>always</u> attractive**! In QED, electrical forces are attractive and repulsive. There are equal number of positive and negative charged virtual particles in the quantum vacuum at any given time because virtual particles are always created in equal and opposite, electrically charged pairs in accordance with the conservation of electric charge. Thus there is always a balance of attractive and repulsive electrical forces in the quantum vacuum, and the quantum vacuum is essentially electrically neutral, overall. If this were not the case, the virtual charged particles of one electrical charge type in the vacuum would dominate over all other physical processes. If the vacuum contained only positively charged virtual particles, there would essentially be a huge cosmological constant problem from the perspective of electrical charge. However, according to general relativity gravity and the principle of equivalence, gravitational force is always attractive. This statement also includes anti-matter. However, if gravity is interpreted as a 'mass charge' in accordance with quantum field theory, than there is no cancellation of the mass charge for the virtual particles of the vacuum. Thus, we are left with the problem of how so much mass in the quantum vacuum amounts to almost no effect on the overall state of the universe.

(2) **QED is formulated in a relativistic, flat 4D space-time with no curvature**. In QED, electrical charge is defined as the fixed rate of emission of photons (strictly speaking, the fixed probability of emitting a photon) from a given charged particle.

Electrical forces are caused by the exchange of photons, which propagate between the charged particles. The photons transfer momentum from the source charge to the destination charge, and travel in flat 4D space-time (assuming no background gravity). From these basic considerations, a successful theory of quantum gravity should have an exchange particle (graviton), which is emitted from a mass particle at a fixed rate as in QED. The 'mass charge' replaces the idea of electrical charge in QED, and the graviton momentum exchanges are now the root cause of gravitational force. Yet, the graviton exchanges must somehow produce disturbances of the normally flat space and time, when originating from a very large mass.

It is common knowledge that mass varies with velocity (special relativity), and one might expect that 'mass charge' must also vary with velocity. However, in QED the electrical force exchange process is governed by a fixed, universal constant ($\alpha$) which is not affected by anything like high speed motion (more will be said about this point later). Should this not be true for graviton exchange in quantum gravity as well? It is strange that gravity, which is also a long-range force, is governed by same form of mathematical law as found in Coulomb's Electrical Force law. Coulomb's Electric Force law states: $F = KQ_1Q_2/R^2$, and Newton's Gravitational Force law: $F=GM_1M_2/R^2$. Therefore, one would expect that these two forces should be very similar. This suggests that there is a deep connection between gravity and electromagnetism. Yet, gravity supposedly has no counterpart to negative electrical charge. Thus, there seems to be no such thing as negative 'mass charge' for gravity, as we find for electrical charge. Furthermore, QED also has no analogous phenomena to the principle of equivalence. Why should gravity be connected with the principle of equivalence, and thus inertia, and yet no analogy of equivalence exists for electromagnetic phenomena? In other words, for electrical forces there exists no connection with accelerated electrical charges as there does for gravitational forces.

In response to the question of negative 'mass charge', *EMQG postulates the existence of negative 'mass charge' for gravity, in close analogy to electromagnetism*. Furthermore, **we claim that negative 'mass charge' is possessed by all anti-particles that carry mass**. Therefore anti-particles, which are opposite in electrical charge to ordinary particles, are also opposite in 'mass charge'. In fact, negative 'mass charge' is not only abundant in nature, it comprises nearly half of all the mass particles (in the form of 'virtual' particles) in the universe! The other half exists as positive 'mass charge', also in the form of virtual particles. Furthermore, all familiar ordinary (real) matter comprises only a vanishing small fraction of the total 'mass charge' in the universe! Real anti-matter seems to be very scarce in nature, and no search for it in the cosmos has revealed it's existence in bulk form to date.

Both positive and negative 'mass charge' appear in huge numbers in the form of virtual particles, which quickly pop in and out of existence in the quantum vacuum, everywhere in empty space. *The existence of negative 'mass charge' is the key to the solution to the famous problem of the cosmological constant* (section 5, ref. 1), which is one of the great unresolved mysteries of modern physics. **Finally, we propose that the negative energy** **or the antimatter solution of the famous Dirac equation of quantum field theory is also the negative 'mass charge' solution of that equation.**

We originally proposed that **the cosmological constant is essentially zero because of the symmetrical production of virtual fermions and anti-fermions particle pairs in the quantum vacuum, which result in a balance of electrical charge and gravitational 'mass charge' in the vacuum**. Thus, the vacuum possess both electrical neutrality and gravitational neutrality. We proposed that there exists a new conservation law for gravitational 'mass charge' (EMQG, ref. 1) so that the creation of virtual, positive and virtual, negative 'mass charge' particle pairs in the quantum vacuum is balanced. At any instant of time, the quantum vacuum has almost exactly equal numbers of positive and negative electrically charged and 'gravitationally charged' fermion particles, which leads to *neutrality of both electrical and gravitational forces in the quantum vacuum, and a cosmological constant of zero.*

However, if this is true, then why did *only* real matter (which posses positive 'mass charge') get created in the early universe? Before we can address this question, we briefly review some of the essential concepts needed from EMQG theory.

### 3. THE VIRTUAL PARTICLES OF THE QUANTUM VACUUM

*Philosophers:    "Nature abhors a vacuum."*

The virtual particles of the quantum vacuum are the key to understanding the cosmological constant. What are virtual particles, and how do we know that they exist in the vacuum? One might insist that the vacuum is completely devoid of everything. The fact is that the vacuum is far from empty. In order to produce a perfect vacuum, one must remove all matter from an enclosure. However, this is still not good enough. One must also lower the temperature down to absolute zero in order to remove all thermal electromagnetic radiation present from the environment. Nernst correctly deduced in 1916 (ref. 8) that empty space is still not completely devoid of all radiation after this is done. He predicted that the vacuum is still permanently filled with an electromagnetic field propagating at the speed of light, called the zero-point fluctuations (sometimes called vacuum fluctuations). This was later confirmed by the full quantum field theory developed in the 1920's and 30's. Later, with the development of QED, it was realized that all quantum fields should contribute to the vacuum state, like virtual electrons and positron particles, for example.

According to modern quantum field theory, the perfect vacuum is teeming with all kinds of activity, as all types of quantum virtual matter particles (and virtual bosons or force particles) from the various quantum fields, appear and disappear spontaneously. These particles are called 'virtual' particles because they result from quantum processes that have small energies and very short lifetimes, and are therefore undetectable.

One way to look at the existence of the quantum vacuum is to consider that quantum theory forbids the absence of motion, as well as the absence of propagating fields (exchange particles). In QED, the quantum vacuum consists of the virtual particle pair creation/annihilation processes (for example, electron-positron pairs), and the zero-point-fluctuation (ZPF) of the electromagnetic field (virtual photons) just discussed. The existence of virtual particles of the quantum vacuum is essential to understanding the famous Casimir effect (ref. 9), an effect predicted theoretically by the Dutch scientist Hendrik Casimir in 1948. The Casimir effect refers to the tiny attractive force that occurs between two neutral metal plates suspended in a vacuum. He predicted theoretically that the force 'F' per unit area 'A' for plate separation D is given by:

$F/A = - \pi^2 h c /(240 D^4)$   Newton's per square meter   (Casimir Force 'F')     (3.1)

The origin of this minute force can be traced to the disruption of the normal quantum vacuum virtual photon distribution between two nearby metallic plates. Certain photon wavelengths (and therefore energies) in the low wavelength range are not allowed between the plates, because these waves do not 'fit'. This creates a negative pressure due to the unequal energy distribution of virtual photons inside the plates as compared to outside the plate region. The pressure imbalance can be visualized as causing the two plates to be drawn together by radiation pressure. Note that even in the vacuum state, virtual photons carry energy and momentum.

Recently, Lamoreaux made (ref. 10) accurate measurements for the first time on the theoretical Casimir force existing between two gold-coated quartz surfaces that were spaced 0.75 micrometers apart. Lamoreaux found a force value of about 1 billionth of a Newton, agreeing with the Casimir theory to within an accuracy of about 5%.

## 4.     GENERAL RELATIVISTIC COSMOLOGICAL CONSTANT PROBLEM

Einstein's introduced a cosmological constant term in his gravitational field equations in 1917 in order to make the universe remain in the steady state (static) or non-expanding state as was believed at that time. Einstein's gravitational field equations describe 4D space-time as a quasi-Riemannian manifold endowed with a metric curvature of the form $ds^2 = g_{ik} dx^i dx^k$, where the metric $g_{ik}$ satisfies the equation given below:

$R_{ik} - (1/2) g_{ik} R = (8\pi G/ c^2 ) T_{ik} - \Lambda g_{ik}$     (4.1)

where, $g_{ik}$ is the metric tensor, $R_{ik}$ is the covariant Riemann curvature tensor. The term $\Lambda$ is the famous cosmological constant. The left-hand side of the above equation is called the Einstein tensor or $G_{ik}$, which is the mathematical statement of space-time curvature that is reference frame independent, and generally covariant. The right hand side $T_{ik}$ is the stress-energy tensor for the entire mass of the universe, G is Newton's gravitational constant, and c the velocity of light. The stress-energy tensor for the vacuum can then be written as:

$$T_{ik\ (vacuum)} = - (\Lambda\ g_{ik}) / (8\pi G / c^2) \tag{4.2}$$

Einstein later abandoned this constant ($\Lambda = 0$) when the astronomer Hubble discovered by astronomical observation that the universe is actually expanding. The cosmological constant can be thought of as the measure of the mass-energy density contained in empty space alone, which would of course be represented by the quantum vacuum virtual particles. In EMQG we originally proposed that the cosmological constant is essentially zero, simply because of the existence of an equal mix of virtual masseon and anti-masseon particles in the quantum vacuum at any given time (ref.1). The masseon particle is the proposed ultimate constituent of all fermion particles. We assumed that the virtual masseon particles must be produced in the vacuum in symmetrical matter and anti-matter masseon pairs according to the principles of quantum field theory. The virtual masseons pairs are oppositely charged in both gravitational 'mass charge' and electrical charge. Approximately equal numbers of positive and negative electrical charge and gravitational mass charge masseons are present in the vacuum at any given time. The symmetrical combination of positive and negative mass-energy yields a net contribution of zero from the quantum vacuum, which is the solution to the cosmological constant problem. It is trivial result to note that the electric charge of the vacuum is neutral as a whole. Before we can probe into the problem of the mass-charge of the vacuum, we review the concept of mass in EMQG.

## 5.     REVIEW OF THE BASIC MASS DEFINITIONS IN EMQG

Since it is the mass contribution from the virtual particles of the quantum vacuum that is the origin of the cosmological constant term, we review the concept of mass in EMQG. We proposed **three *different* mass definitions** for an object in EMQG (ref. 1), which we list below:

(1) **INERTIAL MASS** is the measurable mass 'M' defined in Newton's force law F=MA. This is considered as the absolute mass in EMQG, because it results from force produced by the relative (statistical average) acceleration of the charged virtual particles (masseons) of the quantum vacuum with respect to the charged particles that make up the inertial mass (masseons). To some extent, the virtual particles of the quantum vacuum forms Newton's absolute reference frame. In special relativity this mass definition is equivalent to the rest mass concept of an object.

(2) **GRAVITATIONAL MASS** is the measurable mass involved in the gravitational force as defined in Newton's law $F=GM_1M_2/R^2$. This is the force that is measured on a regular weighing scale. This is also considered as absolute mass in EMQG, and is (almost) exactly the same as inertial mass. The same quantum vacuum process involved in inertia above is also occurring in gravitational mass (section 15.8, ref. 1). However, the roles played by the quantum vacuum particles and the mass particles are reversed. Now it is the virtual particles of the quantum vacuum that is accelerating downwards due to graviton exchanges between the earth and the vacuum, and now it is the mass particles that is at relative rest (with respect to the Earth's surface).

(3) LOW LEVEL **GRAVITATIONAL 'MASS CHARGE'** which is the origin of the pure gravitational force, is defined as the force that results through the exchange of graviton particles between two (or more) quantum fermion particles (ultimately masseons particles). This type of mass is analogous to 'electrical charge', where photon particles are exchanged between electrically charged particles, inducing an electrical force. Note that the forces that arise from gravitational 'mass charge' is very hard to measure directly because it is masked by the background quantum vacuum (electrical force) interactions, which dominates over the graviton force processes. Furthermore, gravitons exchanged between two particles induces an extremely minute gravitational force.

These three forms of mass *are **not** necessarily equal*! It turns out (section 15.8, ref. 1) that the inertial mass is almost exactly the same as gravitational mass, but *not perfectly* equal. All quantum mass particles (fermions, which are made from masseons) have all three mass types defined above. But bosons (only photons and gravitons are considered here) have only the first two mass types. This means that photons and gravitons transfer momentum, and do react to the presence of inertial frames and to gravitational fields, but they do not emit or absorb gravitons. Gravitational fields affect photons, and this is linked to the concept of space-time curvature, described in section 16, ref. 1. It is important to realize that gravitational fields deflect photons, but not by graviton particle exchanges directly with between the earth and the deflected photon. Instead, it is due to an electrical scattering process for the deflected photons, interacting with the falling electrically charged, virtual particles of the vacuum. Similarly, gravitons that travel parallel to the earth's surface are also deflected downwards, again not because of direct graviton exchanges (which would yield a highly non-linear graviton theory). Instead the gravitons also follow a scattering process with the downward accelerating, virtual particles of the vacuum which posses gravitational 'mass charge'.

One might object to his line of reasoning. One might say that if a particle has energy, it must automatically have mass; and if a particle has mass, then it must emit or absorb gravitons. This reasoning is based on Einstein's famous equation $E=mc^2$, which is derived purely from considerations of inertial mass (and later extended for gravitational mass through Einstein's principle of equivalence). In his famous thought experiment, a photon is emitted from a box, causing a recoil to the box in the form of a momentum change, and from this he derives his famous $E=mc^2$. In quantum field theory this momentum change is traceable to a fundamental QED vertex, where a electron (in an atom in the box) emits a photon, and recoils with a momentum equivalent to the photon's momentum '$m_p c$'. We have analyzed Einstein's thought experiment from the perspective of EMQG and concluded that the photon behaves as if it has an effective inertial mass '$m_p$' given by: $m_p = E/c^2$ in Einstein's light box. For simplicity, lets consider a photon that is absorbed by a charged particle like an electron at rest. The photon carries energy and is thus able to do work. When the photon is absorbed by the electron with mass '$m_e$', the electron recoils, because there is a definite momentum transfer to the electron given by $m_e v$, where v is the recoil velocity. The electron momentum gained is equivalent to the effective photon momentum lost by the photon $m_p c$. In other words, the electron momentum '$m_e v_e$' received from the photon when the photon is absorbed is equivalent to the momentum of the photon '$m_p c$', where $m_p$ is the effective photon mass. If this electron later collides with another particle, the same momentum is transferred. The rest mass of the photon is defined as zero. Thus, the effective photon mass is a measurable inertial mass. Note: the recoil of the light box is a backward acceleration of the box, which works against the virtual particles of the quantum vacuum. Thus, when one claims that a photon has a real mass, we are really referring to the photon's ability to impart momentum. This momentum can later do work in a quantum vacuum inertial process.

Does the photon have an effective gravitational mass? By this we mean; does it behave as if it carries a measurable gravitational mass in a gravitational field like the earth (as given by $E/c^2$)? The answer is yes! For example, when the photon moves parallel to the earth's surface, it follows a parabolic curve and deflects downwards. You might guess that this deflection is caused by the graviton exchanges originating from the earth acting on this photon, and that this deflection is the same as that inside an equivalent rocket accelerating at 1g. The amount of deflection is equivalent, but according to Einstein this is a direct result of the space-time curvature near the earth and in the rocket. Our work on the equivalence principle has shown however, that this is not true. The photon deflection is caused by a different reason, but ends up giving the same result. In the rocket, the deflection is simply caused by the accelerated motion of the rocket floor, which carries the observer with it. This causes the observer to perceive a curved path (described as curved space-time). In a gravitational field, however, the deflection of light is real, and caused by the scattering of photons with the downward accelerating virtual particles. The photon scatters with the charged virtual particles of the quantum vacuum, which are accelerating downwards (statistically). The photon moving parallel to the surface of the earth undergoes continual absorption and re-emission by the falling virtual (electrically charged) particles of the quantum vacuum.

The vacuum particles induce a kind of 'Fizeau-like' scattering of the photons (EMQG, ref. 1. Note: this scattering is present in the rocket, but does not lead to photon deflection because only the rocket and observer are accelerated). The photons are scattered because of the electrical interaction of the photons with the falling charged virtual particles of the vacuum. Since the downward acceleration of the quantum vacuum particles is the same as the up-wards acceleration of the floor of the rocket, the amount of photon deflection is equivalent. Under the influence of a gravitational field, photons take on the same downward (component) as the accelerating (charged) virtual particles of the vacuum. This, of course, violates the constancy of the speed of light (section 16, ref. 1). For now, one should note that downward acceleration component is picked up by the photons only during the time the photons are absorbed by the quantum vacuum particles (and thus exist as charged virtual particles). In between virtual particle scattering, the photon motion is still strictly Lorentz invariant, and light velocity is still an absolute constant.

A similar line of reasoning as above applies to the motion of the graviton particle. The graviton has inertial mass because like the photon, it can transmit a momentum to another particle when absorbed in the graviton exchange process during a gravitational force interaction (although considerably weaker then photon exchanges). Like the photon, the graviton deflects when moving parallel to the floor of the rocket (from the perspective of an observer on the floor) and therefore has inertial mass. The graviton also has a gravitational mass (like the photon) when it moves parallel to the earth's surface, where it deflects under the influence of a gravitational field. Again, the graviton deflection is caused by the scattering of the graviton particle with the downward acceleration of the virtual 'mass-charged' particles of the quantum vacuum through an identical 'Fizeau-like' scattering process described above. Unlike the photon however, the scattering is caused by the 'mass charge' interaction (or pure graviton exchanges) of the quantum vacuum virtual particles, and not the electric charge as before. The end result is that the graviton has an effective gravitational mass like the photon. Again a graviton does not exchange gravitons with another nearby graviton, just as a photon does not exchange photons with other photons.

To summarize, in EMQG both the photon and the graviton do **not** carry low level 'mass charge', even though they both possess inertial and gravitational mass. The graviton exchange particle, although responsible for a major part of the gravitational mass process, does not itself carry the property of 'mass charge'. Contrast this to conventional physics, where the photon and the graviton both carry a non-zero mass given by $M=E/C^2$. According to this reasoning, the photon and the graviton both carry mass (since they carry energy), and therefore both must have 'mass charge' and exchange gravitons. In other words, the graviton particle not only participates in the exchange process, it also undergoes further exchanges while it is being exchanged! This is the source of great difficulty for canonical quantum gravity theories, and causes all sorts of mathematical renormalization problems in the corresponding quantum field theory. Furthermore, in gravitational force interactions with photons, the strength of the force (which depends on the number of gravitons exchanged with photon) varies with the energy that the photon carries! In modern physics, we do not distinguish between inertial, gravitational, or low level 'mass charge'. They are assumed to be equivalent, and given a generic name 'mass'. In EMQG, the photon and graviton carry measurable inertial and gravitational mass, but neither particle carries the 'low level mass charge', and therefore do not emit or absorb gravitons.

We must emphasize that gravitons do not interact with each other through force exchanges in EMQG, just as photons do not interact with each other with force exchanges in QED. Imagine if gravitons did interact with other gravitons. One might ask how it is possible for a graviton particle (that always moves at the speed of light) to emit graviton particles that are also moving at the speed of light. For one thing, this violates the principles of special relativity theory. Imagine two gravitons moving in the same direction at the speed of light that are separated by a distance d, with the leading graviton called 'A' and the lagging graviton called 'B'. How can graviton 'B' emit another graviton (also moving at the speed of light) that gets absorbed by graviton 'A' moving at the speed of light? As we have seen, these difficulties are resolved by realizing that there are actually three different types of mass. There is measurable inertial mass and measurable gravitational mass, and low level 'mass charge' that cannot be directly measured. Inertial and gravitational mass have already been discussed and arise from different physical circumstances, but in most cases give identical results. However, the 'low level mass charge' of a particle is defined simply as the force existing between two identical particles due to the exchange of graviton particles only, which are the vector bosons of the gravitational force. Low level mass charge is not directly measurable, because of the complications due to the electrical forces simultaneously present from the quantum vacuum virtual particles.

It would be interesting to speculate what the universe might be like if there were no quantum vacuum virtual particles present. Bearing in mind that the graviton exchange process is almost identical to the photon exchange process, and bearing in mind the complete absence of the electrical component in gravitational interactions, the universe would be a very strange place. We would find that large masses would fall faster than smaller masses, just as a large positive electric charge would 'fall' faster then a small positive charge towards a very large negative charge. There would be no inertia as we know it, and basically no force would be required to accelerate or stop a large mass. Next we must introduce an important new particle of nature.

## 6.      THE PHYSICAL PROPERTIES OF THE MASSEON PARTICLE

In order to understand the principle of equivalence on the quantum level, we had to postulate the existence of a new elementary particle. This particle is the most elementary form of matter or anti-matter, and carries the lowest possible quanta of low level gravitational 'mass charge'. This elementary particle is called the masseon particle (and also comes as anti-masseons, the corresponding anti-particle). The masseon is postulated to be the most elementary mass particle and readily combines with other masseons through a new, hypothetical force coupling which we call the 'primal force'. Presumably, the primal force comes in positive and negative 'primal charge' types. The proposed vector boson mediator of this force is called the 'primon' particle. Since the masseon has not yet been detected through particle collisions, we can safely assume that the primal force is very strong and localized.

It is not necessary to understand the exact nature of the primal force to achieve the important results here. Suffice it to say that the primal force binds together masseon particles to make all the known fermion particles of the standard model. The masseon carries the lowest possible quanta of positive gravitational 'mass charge'. Low level gravitational 'mass charge' is defined as the (probability) fixed rate of emission of graviton particles in close analogy to electric charge in QED. Recall that the graviton is the vector boson of the pure gravitational force. Gravitational 'mass charge' is a fixed constant in EMQG, and is analogous to the fixed electrical charge concept. Gravitational 'mass charge' is **not** governed by the ordinary physical laws of *observable* mass, which appear as 'm' in the various physical theories. This includes Einstein's special relativity theory:

$$E=mc^2 \text{ or } m = m_0 (1 - v^2/c^2)^{-1/2} \tag{6.1}$$

This is why we call it gravitational 'mass charge' or sometimes called the *low-level* mass of a particle, and this should not be confused with the ordinary observable inertial or observable gravitational mass. It will be assumed that when the low level mass is used in this paper, we are talking about the low level gravitational 'mass charge' property of a particle, and the associated graviton exchange process.

Masseons simultaneously carry a positive gravitational 'mass charge', and either a positive or negative electrical charge (defined exactly as in QED). Therefore, masseons also exchange photons with other masseon particles. It is important to note that the graviton exchange process is responsible for the low-level gravitational interaction only, which is not directly accessible to our measurements, and is also masked by the presence of the electrical force component in all gravitational measurements. Masseons are fermions with half integer spin, which behave according to the rules of quantum field theory. Gravitons have a spin of one (not spin two, as is commonly thought), and travel at the speed of light. This paper addresses the gravitational and electromagnetic force interactions only, and the strong and weak nuclear forces are ignored here. Presumably, masseons also carry the strong and weak 'nuclear charge' as well.

Anti-masseons carry the lowest quanta of negative gravitational 'mass charge'. Anti-masseons also carry either positive or negative electrical charge, with electrical charge being defined according to QED. An anti-masseon is always created with ordinary masseon in a particle pair as required by quantum field theory (specifically, the Dirac equation). In EMQG, the anti-masseon is the negative energy solution of the Dirac equation for a fermion, where now the **mass** is taken to be 'negative' as well. Ordinarily, the standard model requires that the mass of any anti-particle is always positive, in order to comply with the principle of equivalence, or $M_i=M_g$. In EMQG, the principle of equivalence is not taken to be an absolute law of nature, and is definitely grossly violated for anti-particles (section 15.4, ref. 1). The anti-particles have positive inertia mass and *negative* gravitational mass, or $M_i=-M_g$.

Thus, a beautiful symmetry exists between EMQG and QED for gravitational and electromagnetic forces. The masseon-graviton interaction becomes almost identical to the electron-photon interaction. There are only two differences between these forces. First, the ratio of the strength of the electromagnetic over the gravitational forces is on the order of $10^{40}$. Secondly, there exists a difference in the nature of attraction and repulsion between positive and negative gravitational 'mass-charges' (as detailed in the table #1 and 2).

In QED, the quantum vacuum as a whole is electrically neutral because the virtual electron and positron (negative electron) particles are always created in particle pairs with equal numbers of positive and negative electrical charge. In EMQG, the quantum vacuum is also gravitationally neutral for the **same** reason. At any given instant of time, there is a 50-50 mixture of virtual gravitational 'mass charges', which are carried by the virtual masseon and anti-masseon pairs. These masseon pairs are created with equal and opposite gravitational 'mass charge'. This is the reason why the cosmological constant is zero (or very close to zero). Half the graviton exchanges between quantum vacuum particles result in attraction, while the other half result in repulsion. To see how this works, we will closely examine how masseons and anti-masseons interact.

The following tables summarize the fundamental electron and masseon force interactions:

**TABLE #1    EMQG MASSEON - ANTI-MASSEON GRAVITON EXCHANGE**

|  | (DESTINATION) | |
|---|---|---|
|  | MASSEON | ANTI-MASSEON |
| (SOURCE) | | |
| MASSEON | attract | attract |
| ANTI-MASSEON | repel | repel |

**TABLE #2    QED ELECTRON - ANTI-ELECTRON PHOTON EXCHANGE**

|  | (DESTINATION) | |
|---|---|---|
|  | ELECTRON | ANTI-ELECTRON |
| (SOURCE) | | |
| ELECTRON | repel | attract |
| ANTI-ELECTRON | attract | repel |

In QED, if the source particle is an electron, it emits photons whose wave function induces repulsion when absorbed by a destination electron, and induces attraction when absorbed by a destination anti-electron. Similarly, if the source is an anti-electron, it emits photons whose wave function induces attraction when absorbed by a destination electron, and induces repulsion when absorbed by a destination anti-electron.

In EMQG, if the source particle is a masseon, it emits gravitons whose wave function induces attraction when absorbed by a destination masseon, and induces attraction when absorbed by a destination anti-masseon. If the source is an anti-masseon, it emits gravitons whose wave function induces repulsion when absorbed by a destination masseon, and induces repulsion when absorbed by a destination anti-masseon. This subtle difference in the nature of graviton exchange process is responsible for some major differences in the way that low-level gravitational 'mass charge' and the electrical charges operate.

It is convenient to think of the photon as occurring in photon and anti-photon varieties (the photon is its own anti-particle). Similarly, the graviton comes in graviton and anti-graviton varieties. Thus, we can say that the masseons emit gravitons, and anti-masseons emit anti-gravitons. The absorption of a graviton by either a masseon or anti-masseon induces attraction. The absorption of an anti-graviton by either a masseon or anti-masseon induces repulsion. Similarly, we can say the electrons emit photons and anti-electrons emit anti-photons. The absorption of a photon by an electron induces repulsion, and the absorption of a photon by an anti-electron induces attraction. The absorption of an anti-photon by an electron induces attraction, and the absorption of an anti-photon by an anti-electron induces repulsion. We now look at the virtual masseons and the quantum vacuum.

## 7.     THE QUANTUM VACUUM AND VIRTUAL MASSEON PARTICLES

Exactly what kinds of virtual particles are present in the quantum vacuum, and how do they contribute to the cosmological constant term? In QED, it is virtual electrons and anti-electrons (and virtual muons and tauons), along with the associated virtual photons. In the standard model of particle physics the quantum vacuum consists of all varieties of virtual fermion and virtual boson particles representing the known virtual matter and virtual force particles, respectively. This includes virtual electrons, virtual quarks, virtual neutrinos for fermions, and virtual photons, virtual gluons, and virtual W and Z bosons for the bosons. In EMQG, we restrict ourselves to the study of gravity and electromagnetism. Therefore, the EMQG quantum vacuum consists of the virtual masseons and virtual anti-masseons, and the associated virtual photons and virtual graviton particles (sometimes, virtual masseon combine to form virtual electrons, etc). Recall that ordinary matter consists only of real masseons bound together in certain combinations to form the familiar elementary particles. We now ask how the virtual electrons/positrons of the QED vacuum behave in the vicinity of a real electrical charge. We want to compare this with virtual masseon and virtual anti-masseon near a large real mass-charge like the earth in our EMQG formulation.

First, we review how the QED quantum vacuum is affected by the introduction of a real negative electrical charge. According to QED, the nearby virtual particle pairs become **polarized** around the central charge. This means that the virtual electrons of the quantum vacuum are repelled away from the central negative charge, while the virtual positrons are attracted towards the central negative charge. Thus for real electrons the vacuum polarization produces charge screening, which reduces the charge of a real electron, when measured over relatively long distances. According to QED, each electron is surrounded by a cloud of virtual particles that winks in and out of existence in pairs lasting a tiny fraction of a second, and this cloud is always present and acts like an electrical shield against the real charge of the electron. Recently, a team of physicists led by D. Koltick of Purdue University in Indiana reported that charge screening of an electron has been observed (ref. 11) experimentally at the KEK collider. They fired high-energy particles at electrons and found that the effect of this cloud of virtual particles was reduced the closer a high-energy particle penetrated towards the electron. They report that the effect of the higher charge for an electron that has been penetrated by particles accelerated to an energy 58 giga-electron volts, was equivalent experimentally to a fine structure constant of 1/129.6. This agreed well with their theoretical prediction of 1/128.5.

Next we study how the EMQG quantum vacuum is affected by the introduction of a large mass. According to EMQG, the quantum vacuum virtual masseon particle pairs are **not** polarized near a large mass, as we found for electrons (as can be seen from table #1 above). The virtual masseon and anti-masseon pairs are *both* attracted towards the mass. This *lack* of polarization which results is the main difference between electromagnetism and gravity. A large gravitational mass (like the earth) does **not** produce vacuum polarization of virtual particles. In gravitational fields, all the virtual masseon and virtual anti-masseon particles of the vacuum have a net average statistical acceleration directed towards a large mass, and produces a net inward (acceleration vectors only) flux of quantum vacuum virtual masseon/anti-masseon particles that can, and do affect other masses placed nearby. In contrast to this, an electrically charged object *does* produce vacuum polarization in QED; where the positive and negative electric charges accelerates towards and away, respectively from the charged object. Hence, there is no energy contribution to other electrical test charges placed nearby (from the vacuum particles only), because the charged vacuum particles contributes equal amounts of force contributions from both directions.

In gravitational fields like the earth, the *lack* of vacuum polarization is responsible for the weak equivalence principle. This is because the electrically charged quantum vacuum masseon/anti-masseon particles can act in unison against a test mass dropped on the earth. Had there been vacuum polarization for masseons, the vacuum particles would act in the two opposite directions, and hence no net vacuum action would result against a test mass. Now we are in a position to study the contribution of the effects of the virtual particles of the quantum vacuum on the cosmological constant.

## 8. THE MASS CONTRIBUTION OF VIRTUAL PARTICLES TO Λ

Before we tackle this problem, let us look at the related question in regards to the *electrical charge* of the vacuum:

WHY IS THE QUANTUM VACUUM *ELECTRICALLY* NEUTRAL?

From pure observation, we would expect that the electrical force contribution from the quantum vacuum to be zero. Yet, the fact remains that there is plenty of electrical charge on the virtual particles of the quantum vacuum. In fact, the electric charge of all the virtual particles of the quantum vacuum represents, by far, the greatest source of electrical charge that exists in the universe at any given instant of time. The electrical charge contributed by the ordinary electrically charged, real matter particle (electrons and protons) makes up an extremely small contribution to the total electrical charge of the universe.

It turns out that the answer to our question is yes, because the vacuum is balanced electrically, so that the sum of all the electrical charges of the quantum vacuum is zero. Let's see why this is so. The virtual bosons do not posses electrical charge, so their contribution is ignored in the analysis. So we will focus on the virtual fermions in the vacuum. In EMQG all virtual fermions are composed of virtual masseon particles which posses the lowest unit of electrical charge and gravitational 'mass-charge' (all masseons are electrically charged). Virtual fermions are constantly created and destroyed in the quantum vacuum. According to the conservation of electrical charge, the virtual fermions are created in particle pairs consisting of matter and anti-matter particle pairs with opposite electrical charge. Therefore, for every negatively charged virtual electron created in the vacuum there is an equal number of positively charged virtual anti-electrons created as well, and the vacuum remains electrically neutral overall. This situation holds as long as the conservation of electrical charge is valid.

A similar argument given above now applies for gravitational 'mass charge' of the quantum vacuum. We now substitute 'mass charge' for electrical charge, where 'mass charge' comes in positive and negative varieties. We use the results from EMQG (ref. 1) that the conservation of gravitational 'mass charge' is valid. The virtual bosons of the quantum vacuum do not posses mass charge, and therefore contribute nothing (section 5, above). We are thus led to the same conclusion in regards to mass charge. However, there is a possible *'loop-hole'* to this argument.

## 9. CONSERVATION OF ELECTRICAL CHARGE AND MASS CHARGE

Are we 100% confident that the conservation of electrical charge and 'mass-charge' holds in every single circumstance? According to EMQG theory, even the tiniest deviation would lead to significant imbalance of virtual matter and virtual anti-matter particles in the quantum vacuum because of the huge numbers of these particles involved. QED (as currently formulated) predicts an infinite number of virtual particles per cubic meter of the vacuum. Other estimates (ref. 1) place an upper cutoff (based on the plank scale being the smallest possible distance), of at least $10^{90}$ virtual particles per cubic meter of the vacuum. Even if only 1 particle creation event in $10^{80}$ violates 'mass charge' conservation than the sum of the vacuum 'mass charge' contribution would be far greater than the number of real fermion particles in the universe, and the quantum vacuum would dominate the dynamics of the universe! Closely related to the conservation of 'mass charge' is the conservation of electrical charge, a subject that is much more well known, and allows us to tie in to the existing standard model of particle physics.

As far as we know there has been **no** confirmed experimental particle events that report any violation of the conservation of electric charge (or any other charge conservation type). However, closely related to conservation of electric charge is the charge conjugation symmetry, denoted by **C.** Charge conjugation means that the physical behavior of a set of interacting particles will be the same if all the particles are replaced by their corresponding antiparticle. For example, a collision between an electron and proton will look *precisely* the same as that of the same collision between a positron (anti-electron) and an antiproton. In other words, the interchange of particle and antiparticle does not modify the dynamics of the interaction. Again, no confirmed violation of **C** symmetry has been reported to date.

There are other symmetries in particle physics, like parity symmetry **P** (or space inversion) and time reversal symmetry **T**. If parity inversion symmetry is violated, then an asymmetry would exist between the real world and a mirror world. If time reversal symmetry is violated a particle interaction would look different if all the motions in the interaction were reversed. It is also possible to define product symmetries, which be obtained by operating two or more of these symmetries simultaneously. For example, if **C** and **P** are operated simultaneously, then we would have the CP symmetry theorem.

Physicists were surprised in 1964 when the CP theorem was violated for neutral kaons. Christenson, Cronin, Fitch, and Turlay (ref. 9) announced that CP was violated for the decay of the neutral kaon particle ($K_L \rightarrow 2\pi$). The CP theorem states that the product of two discrete transformations; Charge Conjugation C and Parity Reflection P is an exact symmetry of nature for all elementary particles. This took the physics community by surprise, because at that time all the symmetry laws were considered as being immutable. Since 1964 the phenomenon of CP violation has attracted great interest from theoretical and the experimental physicists alike. Reference 10 gives a good account of kaon decay physics and CP violation. What is important about CP violation is that it defines an absolute *distinction* between matter and anti-matter, and can be considered a necessary condition for the generation of the baryon asymmetry in our universe.

The baryon asymmetry problem is the name given to the question as to why our universe seems to contain only real matter, and little, if any real anti-matter particles. In particle accelerator experiments, matter seems to be always created in matter / anti-matter particle pairs. It is reasonable to assume that this same process must have happened in the creation of matter in the early universe. Yet our astronomical observations seems to reveal a universe that is entirely composed of real matter, with anti-matter existing in trace amounts. If matter / anti-matter is asymmetrical like the violation of CP symmetry seems to indicate, than it was possible that nature preferred real matter over real anti-matter. One possible scenario is that enormous numbers of matter and anti-matter particles (in unbalanced numbers) were created and mixed in the early universe, and subsequently annihilated each other to form photons. The very slight excess of ordinary matter over antimatter particles forms the universe of stars, clouds, and galaxies we know today. This scenario would also explain the observation that there are many more photons then fermion particles in our universe, most contributing to the cosmological background radiation.

## 10. CONCLUSIONS

Recent high red shift, type I(a) supernovae observations report *an accelerated cosmic expansion,* which implies a small, but positive cosmological constant value. Whether a non-zero cosmological constant turns out to be the explanation for this accelerated cosmic expansion still remains to be seen. We argued that the value of the cosmological constant is zero, provided that the conservation of electrical charge and gravitational 'mass-charge' holds for *every* quantum particle pair creation and annihilation process, without exception. This results in equal numbers of virtual fermion and virtual anti-fermion particles in the quantum vacuum at any given instant of time. Thus the quantum vacuum is electrically neutral and gravitationally neutral, with the attractive gravitational forces of virtual fermion particles exactly balanced by repulsive gravitational forces of virtual anti-fermions. We have acknowledge that negative gravitational force violates the equivalence principle, which is one of the experimentally testable consequences of EMQG theory.

We argued that the virtual bosons do not contribute anything to the cosmological constant, since bosons do not possess 'mass charge' (bosons carry no charge of any kind in EMQG, and are only the charge mediators). Thus, the virtual fermions of the quantum vacuum are strictly the contributors to the value of the cosmological constant. Therefore, the cosmological constant must be very close to zero, as it depends only on the balanced numbers of virtual fermions and virtual anti-fermions in the quantum vacuum. Currently, no violations of the conservation of electric charge have ever been observed, and we assume the same holds for the conservation of 'mass-charge'. However, violations of CP symmetry have been known to occur since the 1960's, which cast doubt on the perfect symmetry between matter and anti-matter.

If matter and anti-matter are perfectly symmetrical, then this leaves the question as to why our universe seems to be composed of only real matter, and very little, if any anti-matter. Therefore the question of the value of the cosmological constant being slightly non-zero can only be fully resolved when a solution to the baryon asymmetry problem for our universe is found.

## 11. REFERENCES

(1) **ELECTROMAGNETIC QUANTUM GRAVITY**: **On the Quantum Principle of Equivalence, Quantum Inertia, and the Meaning of Mass**, by Tom Ostoma and Mike Trushyk, Sept. 11, 1998, LANL E-Print Server, http://xxx.lanl.gov document #: physics/9809042, and the APS E-Print Server.
(2) **CELLULAR AUTOMATA: THEORY AND EXPERIMENT** Edited by H. Gutowitz, 1991. Contains many reprints from Physica D. See pg. 254 for an excellent article by Edward Fredkin.
(3) **SPECIAL RELATIVITY DERIVED FROM CELLULAR AUTOMATA THEORY: The origin of the universal speed limit** by Tom Ostoma and Mike Trushyk, Oct. 7, 1998, LANL E-Print Server, physics/9810010 and the APS E-Print server.
(4) **WHAT HAS AND WHAT HASN'T BEEN DONE WITH CELLULAR AUTOMATA** by by H. V. McIntosh, Nov 10, 1990, LANL Archives.

(5) **INERTIA AS A ZERO-POINT-FIELD LORENTZ FORCE** by B. Haisch, A. Rueda, and H.E. Puthoff; Physical Review A, Feb. 1994. This landmark paper provides the first known proposal that inertia can be understood as the interactions of matter with the surrounding virtual particles.

(6) **THE HIGH RED-SHIFT SUPERNOVA SEARCH - Evidence for a Positive Cosmological Constant** by B. Leibundgut, G. Contardo, et. al, astro-ph/9812042, Dec. 2, 1998.

(7) **THE COSMOLOGICAL CONSTANT IS PROBABLY ZERO** by S.W. Hawking, Physics Letters, Vol. 134B, Num. 6 (1984), pg. 403

(8) **RELATIVITY OF MOTION IN VACUUM** by M. Jaekel, LANL archives, quant-ph/9801071, Jan.30 1998.

(9) **PHYS. REV. LETT. 13 (1964), 138** by Christenson, Cronin, Fitch, and Turlay.

(10) **OVERVIEW OF KAON DECAY PHYSICS** by R. D. Peccei, LANL archives, Hep-ph/9504392, Apr. 27 1995.